\documentstyle[aps,twocolumn,epsfig]{revtex}
\begin{document}
\draft
\date{\today}
\title{The non-extensive version of the Kolmogorov-Sinai entropy at work}
\author{Simone Montangero$^{1}$, Leone Fronzoni$^{1,2}$, 
Paolo Grigolini$^{3,4,5}$}
\address{$^{1}$Dipartimento di Fisica dell'Universit\`{a} di Pisa and INFM,
Via Buonarroti 2, 56127 Pisa, Italy }
\address{$^{2}$Centro Interdisciplinare per lo Studio dei Sistemi Complessi
Via Bonanno 25B, Pisa, Italy }
\address{$^{3}$Dipartimento di Fisica dell'Universit\`{a} di Pisa,
Piazza
Torricelli 2, 56127 Pisa, Italy }
\address{$^{4}$Center for Nonlinear Science, University of North
Texas, P.O. Box 305370, Denton, Texas 76203 }
\address{$^{5}$Istituto di Biofisica del CNR, Via San Lorenzo 26, 
56127 Pisa, Italy }
\maketitle
\begin{abstract}
We address the problem of applying the Kolmogorov-Sinai method of 
entropic analysis, expressed in a generalized non-extensive form, to
the dynamics of the logistic map at the chaotic threshold, which is known 
to be characterized by a power law rather than exponential sensitivity to 
initial conditions. 
The computer treatment is made difficult, if not impossible, 
by the multifractal nature of the natural invariant distribution:
Thus the statistical average is carried out 
on the power index $\beta$. The resulting entropy time evolution
 becomes a smooth and linear
function of time with  
the non-extensive index $Q < 1$ prescribed by the heuristic arguments of 
earlier work,
thereby showing how to make the correct entropic prediction
in the spirit of the
single-trajectory approach of Kolmogorov. 

\end{abstract}


The Kolmogorov-Sinai (KS) entropy \cite{dorfman} is attracting an increasing
interest in the field of chaos since it affords a criterion to establish the
``thermodynamical'' nature of a single trajectory in a way independent of
the observation. This is so because the generating partition
results in a value of the KS entropy independent of the size of the
partition cells \cite{cornfeld}: A fact of fundamental importance to ensure the
objective nature of the KS criterion. There have been in the recent past
some attempts \cite{tsallis1,tsallis2,tsallis3} at generalizing the KS entropy 
by replacing the Shannon entropy, on
which the KS is based, with a non-extensive form of entropy
given by \cite{tsallis}:

\begin{equation}
H_{q}=\frac{1-\sum\limits_{i=1}^{W}p_{i}^{q}}{q-1}.  
\label{entropy}
\end{equation}
The attempts at realizing this purpose are stimulated by 
the exponentially increasing interest
\cite{note} on
this kind of non-extensive entropy.
The parameter $q$, which shall be referred to as \emph{entropix index}, has an
interesting physical meaning. Its departure from the standard value $q=1$,
at which the entropy of Eq. (\ref{entropy}) recovers the traditional Gibbs
structure, signals either the existence of long-range interactions or of
long-time memory. These papers \cite{tsallis1,tsallis2,tsallis3}, however, do
not explicity evaluate the non-extensive version of the KS entropy. 
Rather these authors limit
themselves to assessing numerically the time evolution of the dynamical
property

\begin{equation}
\xi (t)\equiv \lim_{\Delta x(0)\rightarrow 0}\frac{\Delta x(t)}{\Delta x(0)},
\label{sensitivity}
\end{equation}
where $\Delta x(t)$ denotes the distance, at time $t$, between the
trajectory of interest and a very close auxiliary trajectory. The initial
distance between the trajectory of interest and the auxiliary trajectory,
 $\Delta x(0)$, is made smaller and smaller so as to let emerge the kind of
sensitivity of the dynamics under examination. 

The authors of Refs.\cite{tsallis1,tsallis2,tsallis3} related
the function $\xi(t)$ to the non-extensive version of the KS 
entropy with heuristic arguments. With these arguments they
established that the analytical form to assign to $\xi(t)$ for the
non-extensive form of the KS entropy to increase linearly is

\begin{equation}
\xi (t)=[1+(1-Q)\lambda _{Q}t]^{1/(1-Q)},
\label{explicit}
\end{equation}
where $\lambda_{Q}$ is a sort of Lyapunov coefficient.  Throughout this 
letter we shall be referring to  $Q$,
predicted with entropic arguments, as \emph{true entropic index}.
In the specific case $Q<1$, of interest here,  Eq. (\ref{explicit})
means that the distance between the trajectory of interest and the auxiliary
trajectory increases as an algebraic power of time. However, the numerical
calculations made in Refs.\cite{tsallis1,tsallis2,tsallis3} show that the
function $\xi (t)$ \ exhibits wild fluctuations, although the intensity of
these fluctuations fulfills the prediction of Eq. (\ref{explicit}). In Ref. 
\cite{tsallis2} a theoretical prediction was made for the power
index $\beta$, and consequently for the true entropic index
$Q= (\beta-1)/\beta$. This prediction reads:

\begin{equation}
\frac{1}{1-Q}=\frac{1}{\alpha _{\min }}-\frac{1}{\alpha _{\max }},
\label{theory}
\end{equation}
where \ $\alpha _{\min }$\ and \ $\alpha _{\max }$ \ denote the crowding indices
corresponding to the minimum and maximum concentration, respectively.

More recently, the same problem of evaluation of the true entropic
index $Q$, resulting in the linear increase of entropy as a function of
time, was dealt with by the authors of Ref.\cite{latora}, by means of the
numerical calculation of the distribution entropy: The authors 
 of this paper made, in fact, the delicate assumption that if a
true $Q$ exists, making the \emph{trajectory entropy} increase linearly in time,
then the same $Q$ makes the \emph{distribution entropy} increase linear in 
time also. In a sense this letter aims at
checking this important assumption. This is a challenging problem, 
as the ascertainment of the equivalence 
of these two distinct entropy forms is the object of 
discussion also in the case of strong chaos\cite{baranger}

The problem here under study is that of the calculation of

\begin{equation}
H_{q}(N)\equiv \frac{1-\sum_{\omega_{0}...\omega_{N-1}}
p(\omega_{0}...\omega_{N-1})^{q}}{q-1}, 
\label{kolmogorov}
\end{equation}
where $p(\omega _{0}...\omega _{N-1})$ is the probability of finding the
cylinder corresponding to the sequence of symbols $\omega _{0}...\omega
_{N-1}$\cite{beck}. This entropy expression affords a rigorous way of 
defining the earlier 
introduced concept of \emph{true entropic index}. If it exists, $Q$ is the 
value of the entropic index $q$ making the entropy of Eq.(\ref{kolmogorov}) 
increase
linearly in time. In the case $q = Q = 1$ the $\lim_{N \rightarrow \infty}  
H_{q}(N)/N$ becomes the ordinary Kolmogorov-Sinai (KS) entropy.
In an earlier work\cite{luigi}
a numerical calculation was made to establish $Q$ in the case of a text 
of only two symbols, with strong correlations. 
The case of many more symbols would be beyond the range
of the current generation of computers.
However, when the sequence of symbols is generated by dynamics, as in 
the case here under study, and the function $\xi(t,x)$ of 
Eq.(\ref{sensitivity}) is available (for convenience, we make now 
explicit the dependence on the initial condition $x$), it is possible 
to
adopt the prescription of Ref.\cite{jin}, which writes $H_{q}(t)$ of 
Eq.(\ref{kolmogorov}) as

\begin{equation}
H_{q}(t)\equiv \frac{1-\delta ^{q-1}\int dxp(x)^{q}\xi (t,x)^{1-q}}{q-1},
\label{jin}
\end{equation}
where the symbol $t$ denotes time regarded as a continuous variable. 
In fact, when
the condition $N>>1$ applies, it is legitimate to identify $N$ with $t$. The
function $p(x)$ denotes the equilibrium distribution density and $\delta $
the size of the partition cells: According to Ref.\cite{jin} the phase
space, a one-dimensional interval, has been divided into $ W = 1/\delta $ 
cells of
equal size. 

There is now an important remark to make: The 
non-extensive form of KS entropy should read as follows,
\begin{equation}
h_{Q} = \delta ^{Q-1} \lim_{t \rightarrow \infty} 
\frac{1}{t}\frac{\int dxp(x)^{Q}\xi (t,x)^{1 -Q}}{1-Q}, 
Q \neq 1.
\label{KSgeneralized}
\end{equation}
This apparently means that  
leaving the ordinary condition $Q=1$  has 
the unwanted effect of making the generalized form of KS entropy 
dependent on $\delta$, thereby losing what we consider to be the most 
attractive aspect of the KS entropy. We are inclined to believe
that this apparent weakness is, on the contrary, an element of 
strength. We shall see that in the case under discussion in this 
letter, due to the multifractal character of the natural invariant 
distribution $p(x)$, the prescription of Eq.(\ref{KSgeneralized})
results in a rate of entropy increase independent of the cell size. 
In the case of the Manneville map\cite{manneville}, 
$x_{n+1}= x_{n}+ x_{n}^{z},\; mod \; 1$, it is shown\cite{anna} 
that in the range $3/2 < z < 2$ the  
stationary correlation 
function of the variable $x$ exists and it is not integrable, thereby
suggesting a possible breakdown of the ordinary KS entropy. However,
in this case the natural invariant distribution is smooth, rather than  multifractal, 
and as a consequence of that, the request of the independence of
the cell size and the adoption of the prescription of Eq.(\ref{jin})
yield the condition $Q=1$, in agreement with the conclusions
of the work of Ref.\cite{gaspard}. In the case $Q=1$  it is
straightforward to prove that Eq.(\ref{jin}) results in

\begin{equation}
H_{1}(t)=\int dxp(x)\ln \xi (t,x).  \label{normalcase}
\end{equation}
On the other hand, in the ordinary case Eq.(\ref{explicit}) becomes 
\begin{equation}
\xi (t,x)=\exp (\lambda (x)t),  \label{exponential}
\end{equation}
thereby making Eq.(\ref{normalcase}) result in the well known Pesin relation
\cite{pesin}
\begin{equation}
h_{KS}\equiv \lim_{t\rightarrow \infty }\frac{H_1(t)}{t}=\int dxp(x)\lambda
(x).
\label{pesin}
\end{equation}
In conclusion, in the case of a smooth invariant distribution, either
strongly or weakly chaotic,
 the prescription of Eq.(\ref{jin}) 
coincides with the Pesin theorem, which allows us to replace 
the direct calculation of the KS entropy with the numerically easier
problem of evaluating Lyapunov coefficients. The case of fractal 
dynamics implies the existence of the true $Q \neq 1$, which has to be
properly detected looking for the value of $q$ making $H_{q}(t)$ 
linearly dependent on $t$. This letter is devoted to providing the 
guidelines for this search.

We shall focus our attention on
the calculation of

\begin{equation}
\Xi_{q}(t,\delta)\equiv \delta ^{q-1}\int dxp(x)^{q}\xi (t,x)^{1-q}.  
\label{meandep}
\end{equation}
Note that if $\xi(t)$ of Eq.(\ref{explicit}) depended on $x$, the value 
$\Xi_{Q}^{1/(1-Q)}$ resulting from the joint use of  Eqs.(\ref{meandep})
 and (\ref{explicit}) 
would afford a simple recipe to determine the statistical average 
 $\langle \lambda_{Q}(x) \rangle$ at $q = Q$. In principle, 
$\Xi_{q} (t,\delta )$ depends on the cell size $\delta$. However, we plan
to prove that if it is properly evaluated, 
this quantity turns out to be independent of $\delta $. 
As done in the earlier work of 
Refs.\cite{tsallis1,tsallis2,tsallis3,latora}, we study the logistic 
map:
\begin{equation}
x_{n+1} = 1 - \mu |x_{n}|^{2}, x \in [-1,1]
\label{logistic}
\end{equation}
with the control parameter $\mu = 1.4011551\ldots$, namely, at the 
threshold of transition to chaos.  In this case the invariant
distribution is multifractal, and consequently, expressed as a function 
of $x$,
looks like a set of sharp peaks, which, in turn, through repeated zooming, 
reveal to consists of infinitely many other, sharper, peaks.
This means that the direct evaluation
of Eq.(\ref{jin}) is hard, since it is difficult 
to ensure numerically that these fractal properties are reproduced
at any arbitrarily small spatial scale. We have to look for a 
different approach.

Let us replace the average over $x$ with  the average over the crowding
power index $\alpha $. In the long-time limit we obtain\cite{beck}

\begin{equation}
\Xi_{q} (t)\equiv \delta ^{q-1}\int d\alpha \delta ^{q\alpha -f(\alpha
)}t^{\beta (t,\alpha )(1-q)}.  \label{averagingoncrowdingindex}
\end{equation}
This equation rests on assuming dependence on the initial condition  only 
through
the power law index $\beta (t,\alpha )$ itself which, in fact, according
to Anania and Politi Ref.\cite{anania}, reads

\begin{equation}
\beta (t,\alpha )=\frac{1}{\alpha (t)}-\frac{1}{\alpha }.  \label{anania}
\end{equation}
We shall show with theoretical and numerical arguments
that this relation yields  Eq.(\ref{theory}) for the exact value of 
$Q$.
The symbol $\alpha $ denotes the crowding index corresponding to a given
initial condition $x,$ namely the position of the trajectory at 
$t=0$, and
the symbol $\alpha (t)$ denotes the crowding index corresponding to the
position of the same trajectory at a later time $t>0$. According to 
Anania and Politi \cite{anania}
\begin{equation}
\alpha(t)=\frac{ln(1/t)}{ln|x(t+2^{k}) -x(t)|}, 
\label{tesi}
\end{equation}
where $k$ indicates the $k$-th generation of the Feigenbaum
attractor.

Before proceeding, let us make an assumption which has the effect of 
accomplishing, within the non-extensive perspective, the
Kolmogorov program of an entropy independent of the  size of the partition cells. 
First of all, let us rewrite Eq. (\ref{averagingoncrowdingindex}) in 
the following equivalent form:
\begin{equation}
\Xi_{q} (t) = \int d\alpha \ e^{W(f(\alpha) - q\alpha + q - 1)}
e^{V(q-1)\beta(\alpha,V)},
 \label{crossingstep}
\end{equation}
where:
\begin{equation}
W \equiv - ln \delta
\label{cell}
 \end{equation}
 and
\begin{equation}
V \equiv - ln (1/t).
\label{time}
 \end{equation}
 Let assume now:
 \begin{equation}
 W << V.
 \label{keyassumption}
 \end{equation}
 Under the plausible condition that the functions $f(\alpha) - q\alpha + q - 1$ 
 and $\beta(\alpha,V)$ are not divergent,
 this assumption has the nice effect of producing
 \begin{equation}
\Xi_{q} (t) = \int d\alpha 
\ e^{V(q-1)\beta(\alpha,V)}.
\label{cornfeld}
\end{equation}
At this stage we make another crucial step. This is suggested by the 
work of Hata, Horita and Mori\cite{hata}. The idea is that of using
$\beta(\alpha,V)$ as independent variable so as to write 
Eq.(\ref{cornfeld}) either as:
\begin{equation}
\Xi_{q} (V) = \int d\beta P(\beta,V) 
e^{V(q-1)\beta(\alpha,V)}.
 \label{V}
\end{equation}
or under the equivalent form:
\begin{equation}
\Xi_{q} (t) = \int d\beta P(\beta,t) 
t^{\beta(q-1)}.
 \label{t}
\end{equation}

We follow Ref.\cite{hata} again and we adopt the asymptotic 
property\cite{ellis}:
\begin{equation}
P(\beta,t) = t^{-\psi(\beta)}P(\beta,0) .
 \label{ellis}
\end{equation}
The numerical calculation of the function $\psi(\beta)$ is done with a 
criterion different from that adopted in Ref.\cite{hata}. The authors
of Ref.\cite{hata} fix a window of a given size $t$ and move it along 
the sequence for the purpose of evaluating the frequency of presence 
within this window of a given algebraic index $\beta$. This means that 
they make an average over many different initial conditions. We, on 
the contrary, fix a given initial condition, and we increase the size 
of the window, the left border of which coincides with the initial 
time condition $t=0$, whereas the right border, at the distance $t$ from the former, 
runs over the whole range of observation times. This different criterion 
is dictated by the specific purpose of evaluating the quantity of 
Eq.(\ref{t}) in a way compatible with using only one single trajectory, while 
apparently the authors of Ref.\cite{hata} do not feel the need of 
fitting this constraint. 
 
\begin{figure}[t]
\begin{center}
\begin{picture}(100,120)
\put(-48,-3){\epsfig{figure=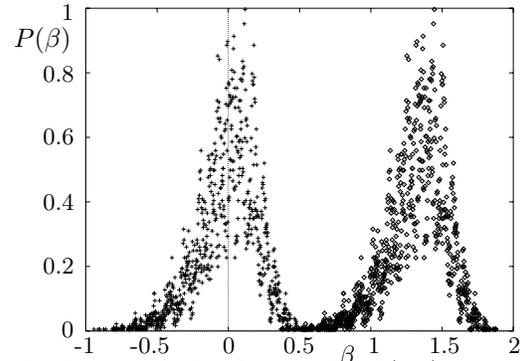,width=6.5 cm}}
\put(-58,110){{$P(\beta)$}}
\put(65,-10){$\beta$}
\put(-40,0){\small{0}}
\put(-47,24){\small{0.2}}
\put(-47,48){\small{0.4}}
\put(-47,73){\small{0.6}}
\put(-47,97){\small{0.8}}
\put(-40,120){\small{1}}
\put(-35,-6){\small{-1}}
\put(-14,-6){\small{-0.5}}
\put(21,-6){\small{0}}
\put(44,-6){\small{0.5}}
\put(76,-6){\small{1}}
\put(99,-6){\small{1.5}}
\put(129,-6){\small{2}}
\end{picture} 
\end{center}
\caption{The distribution density $P(\beta,t)$ as a function of $\beta$.
The distribution on the left has been evaluated using as initial 
condition $\alpha_{min}$; the distribution on the right uses as 
initial condition $\alpha_{max}$. The time $t$ has been set $2^{15} < t < 2^{18}$.}
\end{figure}

The initial condition chosen is $x=1$.
The reason for this choice is widely 
discussed in Refs.\cite{tsallis1,tsallis2,tsallis3}. This is so because
$\alpha(x =1) = \alpha_{min}$, therefore ensuring the condition of 
maximum expansion.
 In Fig. 1 we show that the choice of
 $\alpha_{min}$ rather than $\alpha_{max}$ shifts
 the distribution $P(\beta,n)$ from the right to the left, namely 
 from a condition close to that of Ref.\cite{hata}, to
 a condition favorable for the emergence of $Q$. 
  We note that the $\beta$-distribution
 does not drop to zero beyond the value $\beta \approx 1.3$, which, 
 according to Eq.(\ref{anania}) is the maximum possible value of the 
 power index $\beta$. This is a consequence of the fact that the 
 theoretical prescription of Eq.(\ref{tesi}) refers to the time 
 asymptotic limit $t \rightarrow \infty$, whereas the numerical calculation is carried 
 out with an upper bound on time: The maximum value of time explored 
 is in fact $t_{max} = 2^{18}$. It is expected that with the increase
 of the time upper bound the distribution tends to drop to zero for 
 values of $\beta$ larger than the maximum possible value.

 The result of the corresponding numerical calculation is shown in 
 Fig. 2. To make more evident that the central 
 curve, with $q \approx 0.25$, is that corresponding to the true $Q$,
 we adopt the same procedure as that used by the authors of 
 Ref.\cite{latora}. We have fitted the curves $H_{q}(t)$ of Fig. 3 in 
 the interval $[t_{1},t_{2}]$ with the polynomial $H(t) = a + bt + ct^{2}$.
 We define $R=|c|/b$ as a measure of the deviation from 
 the straight line. We expect that $q=Q$ results in $R=0$. We choose
 $t_{1}= 100$ and $t_{2}= 1000$ for all $q$'s. 
 In the insert of Fig.2 we show that $R$ becomes virtually equal to zero
 for $q = 0.25$, which is very close to the value $q=0.24$ found by the authors 
of Ref.\cite{latora}.
 
 The shift from the left to the right distribution shown in Fig.1 
 is of fundamental importance to find the correct $Q$. Further 
 evidence of this fact is obtained by using analytical arguments to
 evaluate the integral of Eq.(\ref{t}).
We proceed as follows. First, we fit the numerical
data on $\psi(\beta)$
 by means of the function
\begin{equation}
\psi_{fit}(\beta)= const \cdot (\beta - \beta^*)^2,
\label{parabola}
\end{equation}
with $\beta^*= 1.35 \pm 0.05$. For
 $t \rightarrow \infty$ the resulting $P(\beta)$ becomes equivalent to a 
 Dirac  $\delta$ function centered at $\beta^*= 1.35 \pm 0.05$,
thereby making straightforward to find
\begin{equation}
\Xi_{q}(t) \propto \frac{ t ^{\beta^* (1-q)}}{\ln t }.
\end{equation}
If we neglect the logarithm term, the entropy growth
becomes linear in time at $Q =0.259\dots \pm 0.02$. 
A more accurate fitting procedure, taking the distribution asymmetry
into account, was proved to produce virtually the same result.
This provides a strong support to the 
conclusion of the
numerical results of Fig. 2. As already mentioned earlier, we expect that
with increasing the time upper bound the 
$\beta$-distribution will reach a maximum at $\beta = 1/\alpha_{min} - 
1/\alpha_{max}$ and will drop abrutly to zero for larger $\beta$.
Consequently, the resulting true $Q$ is expected to coincide with that 
of Eq.(\ref{theory}). In conclusion, we are convinced that the slight 
discrepancy between our numerical result and the theoretical prediction 
of Refs.\cite{tsallis1,tsallis2,tsallis3} is essentially due to the 
fact that the numerical observation is carried out at finite times.

\begin{figure}[ht]
\begin{center}
\begin{picture}(100,120)
\put(-48,-3){\epsfig{figure=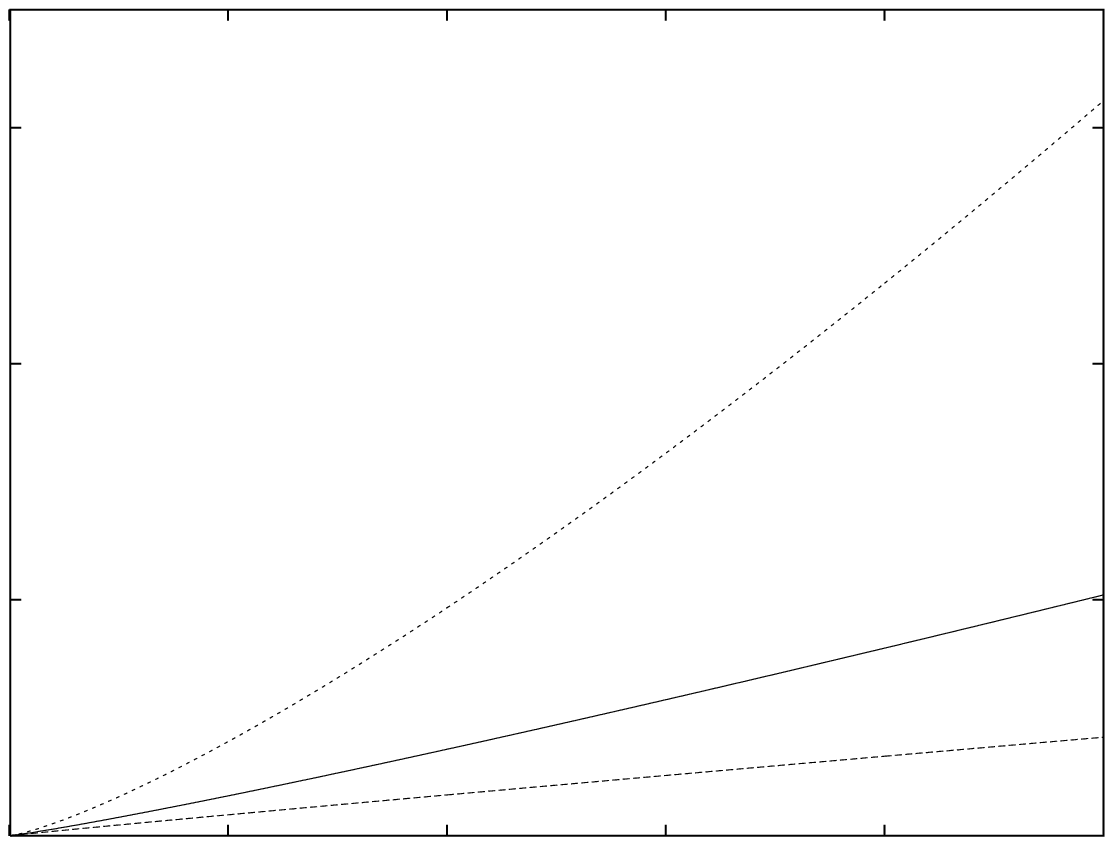,width=6.5 cm}}
\put(-58,115){{$H_q(t)$}}
\put(49,-10){$t$}
\put(-38,-5){\small{0}}
\put(-52,34){\small{2000}}
\put(-52,68){\small{4000}}
\put(-52,102){\small{6000}}
\put(-5,-5){\small{200}}
\put(27,-5){\small{400}}
\put(60,-5){\small{600}}
\put(92,-5){\small{800}}
\put(122,-5){\small{1000}}
\end{picture} 
\put(-123,65){\epsfig{figure=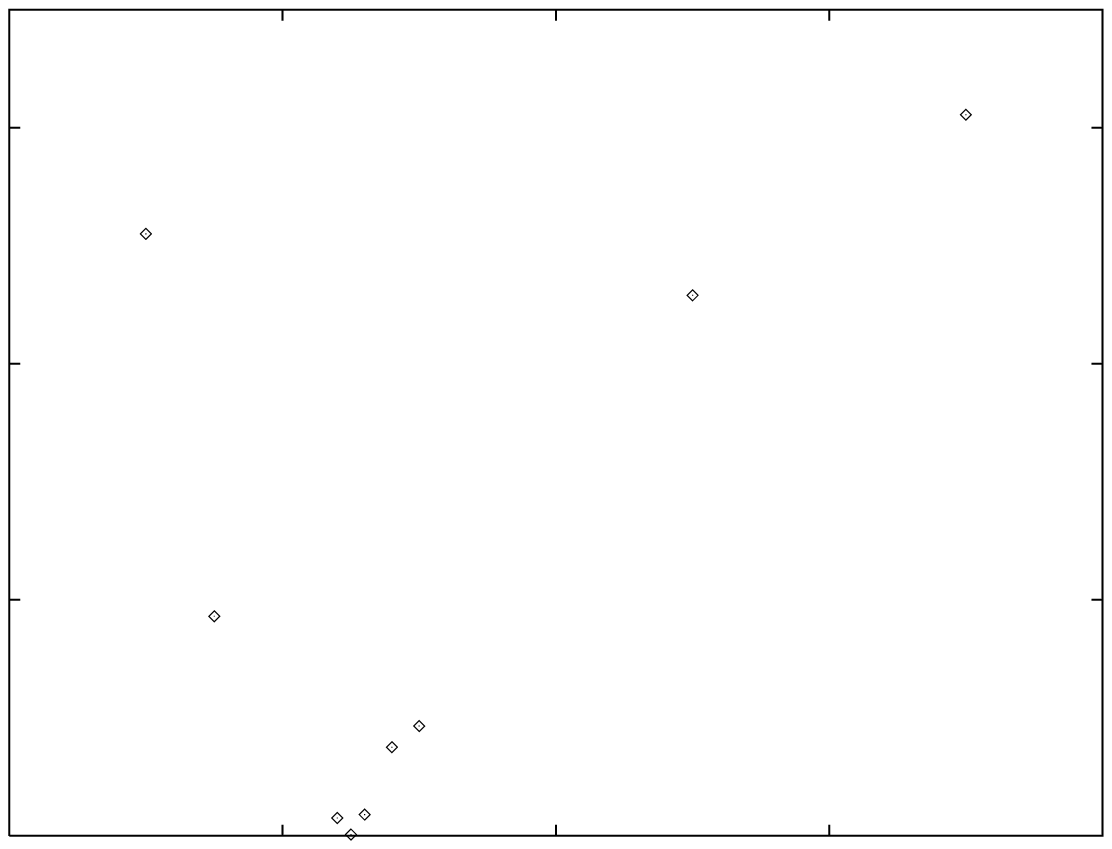,width=2.8 cm}}
\put(-80,55){\small{$q$}}
\put(-128,118){\small{$R$}}
\put(-120,110){\tiny{$6$}}
\put(-120,94){\tiny{$4$}}
\put(-120,78){\tiny{$2$}}
\put(-120,62){\tiny{$0$}}
\put(-103,62){\tiny{$0.2$}}
\put(-85,62){\tiny{$0.4$}}
\put(-67,62){\tiny{$0.6$}}
\put(-50,62){\tiny{$0.8$}}
\end{center}
\caption{ The function $H_{q}(t)$ of Eq.(\ref{jin}) as a function of time for 
three different values of the entropic index $q$. The values of $q$, 
from the bottom to the top curve, are $q = 0.35, 0.25, 0.15$. The insert 
shows $R$ vs $q$. The values of $R$ have been multiplied by $10^3$.}
\end{figure}

The 
importance of this paper goes much beyond cheking the prediction of Ref. 
\cite{latora}. The result obtained is equivalent to observing only one 
trajectory moving from the initial condition $x=1$. This 
 central result is
made possible by the use of the average over $\beta$ suggested by the 
important work of Hata et al.\cite{hata} as well as by the  
result of Ref.\cite{jin}. The method of Ref.\cite{jin} 
proves to be an efficient way of 
expressing the dependence of the trajectory entropy on the sensitivity to the initial conditions. We are convinced that Eq.(\ref{jin}) can be regarded as the 
proper non-extensive generalization of the Pesin theorem. Consequently, 
under the  assumption that the dependence on initial conditions is realized 
only through $\beta(t,\alpha)$,  the slope of the curve of Fig. 2, 
corresponding to $q = Q =0.25$, is the genuine extensive KS entropy of 
this archetypical condition of weak chaos.

We thank C. Tsallis for reading the draft of this paper and 
illuminating suggestions.

\end{document}